\documentclass[10pt, journal]{IEEEtran}

\usepackage{blindtext}
\usepackage{amsmath}
\usepackage{amssymb}
\usepackage[utf8]{inputenc}
\usepackage[textsize=tiny]{todonotes}

\usepackage{graphicx}
\usepackage{subfig}
\usepackage{circuitikz}
\usepackage{tikz}
\usepackage{footnote}
\usepackage{booktabs}
\usepackage{multirow}
\usepackage{cite}
\usepackage{url}
\usepackage{array}
\usepackage{rotating}

\usepackage{tikz}
\usetikzlibrary{circuits.logic.US,circuits.logic.IEC}
\usetikzlibrary{matrix}
\usetikzlibrary{backgrounds,calc,decorations.pathreplacing,matrix}
\usetikzlibrary{circuits,shapes}
\usetikzlibrary{arrows}
\tikzstyle{circ}=[circle, draw=black, font=\small\sffamily\bfseries,inner sep=6pt]
\tikzstyle{trigl}=[
isosceles triangle,  
draw,
shape border rotate=90,
inner sep=3,
font=\small\sffamily\bfseries,
isosceles triangle apex angle=60,
isosceles triangle stretches]

\newtheorem{example}{Example}

\begin{document}
	
	\title{Near Zero-Energy Computation Using Quantum-dot Cellular Automata}	
	\author{Frank~Sill~Torres, Philipp~Niemann, Robert~Wille,  and~Rolf~Drechsler}

\maketitle

\begin{abstract} 
Near zero-energy computing describes the concept of executing logic operations below the $(k_BT \ln 2)$~energy limit. Landauer discussed that it is impossible to break this limit as long as the computations are performed in the conventional, \mbox{non-reversible} way. 
But even if reversible computations were performed, the basic energy needed for operating circuits realized in conventional technologies is still far above the $(k_BT \ln 2)$~energy limit, i.e.\, 
the circuits do not operate physically reversible. 
In contrast, novel nanotechnologies like \mbox{\emph{Quantum-dot Cellular Automata}} (QCA) allow for computations with very low energy dissipation and, hence, are promising candidates for breaking this limit.
Accordingly, the design of reversible QCA circuits is an active field of research. 
But whether QCA in general (and the proposed circuits in particular) are indeed  able to operate in a logically \emph{and} physically reversible fashion is unknown thus far, because neither physical realizations nor appropriate simulation approaches were available yet.
In this work, we address this gap by utilizing an established theoretical model that has been implemented in a physics simulator enabling a precise consideration of how energy is dissipated in QCA designs.
Our results provide strong evidence that QCA is indeed a suitable technology for near zero-energy computing. 
Further, the first design of a logically and physically reversible adder circuit is presented which serves as 
proof-of-concept for future circuits with the ability of near zero-energy computing.

\end{abstract}


\section{Introduction}\label{sec:intro}
In 1961, Rolf Landauer argued that any non-reversible computational process, i.\,e., each computation in which information is lost, results in a minimum energy dissipation of $(k_BT \ln 2)$ per bit erased (with $k_B$ being the Boltzmann constant and $T$ being the temperature of the system's thermal environment~\cite{Landauer61}). The validity of this thermodynamic limit in computation, also known as \emph{Landauer's principle}, has been disputed ever since. Recently, it eventually got experimentally verified in several technologies---confirming that there is a physical limit in non-reversible computation~\cite{Berut12,Hong16,Orlov12,Neri16}. 

This significantly affects conventional computation technologies, since they mostly rely on non-reversible operations such as NAND, which non-reversibly transform two input bits into a single output bit and, hence, lose one bit of information. 
Although the energy dissipation caused by such information losses (which usually are much larger than Landauer's lower bound~\cite{Gershenfeld:1996})  was considered negligible for a long time, the miniaturization of computational devices as well as improved material and fabrication processes led to energy dissipations of today's circuits and systems which come quite close to \emph{Landauer's limit} constituted by~$(k_BT \ln 2)$ per bit erased. 
In other words, in order to continue the \mbox{ever-increasing} reduction of energy consumption as observed in the past decades, and thus, reach the ability of near zero-energy computing, complementarily different ways of computation are required, i.\,e.~computations which can conduct operations without losing information.

Reversible computations are an obvious alternative. Here, all computations are realized through \emph{reversible} operations, i.\,e.~bijective functions which map each input pattern to a unique output pattern.  Already in 1973, Bennett proved in his seminal work that energy dissipation is reduced or even (theoretically) eliminated if computations are \mbox{information-lossless}~\cite{Bennett73}. In fact, he showed that any circuit and system with a (theoretical) energy dissipation of zero must rely on reversible computation. 
This motivated entire fields of research such as adiabatic computations (see e.\,g.~\cite{Zulehner19}), reversible energy recovery logic (see e.\,g.~\cite{ye1996energy}), or the design and realization of corresponding reversible circuits (see e.\,g.~\cite{zulehner2017one,Chaves19}).
Today, it is seen as a fact that future computation technologies have to be reversible in order to overcome limitations due to energy dissipation~\cite{Frank17}.
Accordingly, technologies are desired which come with \mbox{low-energy} dissipation, i.\,e. with energy dissipation close to the $(k_BT \ln 2)$ energy limit, and, at the same time, can be tweaked to conduct operations in a reversible fashion---eventually allowing for near zero-energy computing~\cite{DeBenedictis16}.

In this regard, \mbox{\emph{Quantum-dot Cellular Automata}}~(QCA)~\cite{Lent97} are a promising candidate. They employ a \emph{Field-Coupled Nanotechnology}~(FCN) in which information is stored in terms of the polarity of small cells and can be propagated to adjacent cells using electrostatic force (Coulomb interaction). This allows for representing and processing information with remarkably low energy dissipation (see e.\,g.~\cite{Timler02,Pitters11}) and makes QCA an interesting candidate for near zero-energy computing.
	
This motivated the consideration of several reversible QCA designs (see e.\,g.~\cite{Chaves15,Singh17,Mukhopadhyay15,Chabi17,Sen17}). 
But all of these previous works considered only the logic level when making the desired function reversible, i.\,e. the proposed designs indeed realize bijections, while they did not address information loss at the gate level (physical level) and either provide no physical realization at all or employ non-reversible gates. 
This is obviously not enough to enable near zero-energy computing as it does not address the conceptual problem. Instead,  near zero-energy computing is only feasible if reversibility is kept until the physical level and if energy dissipation remains below $(k_BT \ln 2)$ per operation~\cite{DeBenedictis16}. 
In this regard, specific clocking strategies as proposed in~\cite{Lent06,Ottavi11} seem much more promising. However, due to the absence of a method for exact estimation of the energy dissipation in QCA designs, the assumptions could not be validated. 
This restriction holds for all existing proposals, as available simulation approaches do not consider energy dissipation at all (e.\,g.~\cite{Walus06}) or restrict themselves to worst-case scenarios (e.\,g.~\cite{Bhanja09}) such that they do not allow for a precise estimation of the energy dissipation in QCA designs. 
As a consequence, it is not known yet whether QCA designs indeed allow for implementing reversibility on logical \emph{and} physical level, and thus, enable operations with near zero-energy dissipation.


In this work, we are going to change this \mbox{state-of-the-art} by delivering strong evidence for this conjecture.
To this end, we utilize an established theoretical model that has been implemented in a physics simulator enabling a precise consideration of how energy is dissipated in QCA designs~\cite{Timler02,Bhanja09,Sill18}. 
For the first time, we are using this approach to explore the capability of QCA designs to operate with near zero-energy consumption.

The results clearly indicate that:
\begin{itemize}
\item Basic QCA building blocks such as wires and primitive logic gates can indeed be operated in a logically \emph{and} physically reversible way, and thus, with an energy dissipation below $(k_BT \ln 2)$ per operation.
\item The same also holds for more complex QCA designs made up of these reversible building blocks---thereby demonstrating QCA's general capability to conduct computations with an energy dissipation below the crucial $(k_BT \ln 2)$ energy barrier (and without significant architectural changes such as used in~\cite{Lent06,Ottavi11}). 
\item Design methodologies for QCA as proposed in~\cite{Huang05} or~\cite{Walter19a} can be applied for composing reversible circuits that operate with near zero-energy dissipation.
\end{itemize}

The remainder of this work is structured as follows: Section~\ref{sec:energy_model} reviews the basics on QCA as well as their energy dissipation. 
Afterwards, Section~\ref{sec:bb} analyses how energy is dissipated by primitive QCA building blocks such as wires or functional gates (like OR) and discusses how these can be operated in logically and physically reversible way.
Based on that, a logically and physically reversible adder circuit is introduced and the results of energy simulations for larger QCA designs are discussed in Section~\ref{sec:results}---showing that QCA indeed allow for near zero-energy computing.
Finally, the paper is concluded in Section~\ref{sec:concl}.


\section{Background}
\label{sec:energy_model}

In this work, we aim for validating the capability of the QCA technology to operate with near zero-energy dissipation.
Since corresponding physical realizations are not available yet, we rely on simulations instead. 
Therefore, we rest on an established theoretical model which has been integrated in a physics simulator~\cite{Timler02,Bhanja09,Sill18}, enabling the precise analysis of energy transfers in QCA. 
This section reviews the basic concepts of QCA and the physical model for the energy transfer in QCA, and introduces the resulting tool.

\subsection{Quantum-dot Cellular Automata}\label{sec:basics_QCA}

The term \emph{Quantum-dot Cellular Automata}~(QCA) refers to an emerging, field-coupled nanotechnology which takes an alternative approach to processing information and performing computations that is fundamentally different from today’s established technologies.
In fact, information is stored in terms of the polarity of small, square-shaped cells that are arranged in a \emph{grid} structure. Within a cyclic process, a cell's information is regularly erased and a new polarization is determined based on the polarization of the surrounding cells, i.\,e. by electrostatic force (Coulomb interaction).

\begin{figure}
	\centering
	\subfloat[States in a QCA\label{fig:qca_states}]{\hspace{3mm}\includegraphics[scale=1]{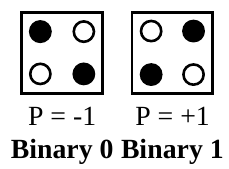}\hspace{3mm}} \hfil
	\subfloat[QCA Majority\label{fig:qca_maj}]{\hspace{3mm}\includegraphics[scale=1]{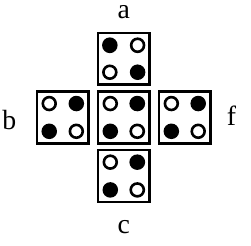}\hspace{3mm}}
	\caption{QCA states and Majority operation}\label{fig:qca_basics}
\end{figure}

More precisely, a QCA \emph{cell} is typically composed of (1)~four quantum dots situated at the corners of the cell as well as (2)~two free and mobile electrons which are able to tunnel between adjacent dots~\cite{Anderson14}.  The electrons may not tunnel to the outside of the cell due to a potential limit, and also tunneling within the cell can temporarily be prevented---thereby forcing the electrons to remain stationary at one of the quantum dots and, thus, leading to a stable state.  For an illustration of QCA cells with stationary electrons, see Fig.~\ref{fig:qca_basics} where the electrons are represented by black dots.

Electrons experience mutual repulsion due to Coulomb interaction. As a consequence, they tend to locate themselves as far as possible from each other when the intracellular tunneling is prevented and they are being forced to become stationary.
Consequently, an isolated cell will assume either of two stable energy states in which the electrostatic forces are minimal. These states are termed \emph{cell polarizations} and are usually denoted as $P = -1$ and $P = +1$.
The described behavior allows for an encoding of binary information by associating each polarization with a binary value. To this end, one usually identifies $P = -1$ with a binary 0 and $P = +1$ with a binary 1 as shown in Fig.~\ref{fig:qca_states}.

Moreover, when multiple cells are placed close to each other, the polarization of each cell is influenced by the polarization of the others. More precisely, the mutual repulsion causes electrons to avoid a quantum dot if the neighboring quantum dots of adjacent cells are populated by other electrons. 
This effect can be exploited for the realization of circuit elements, such as the binary OR and AND that can be derived as special cases of the 3-input \emph{majority gate} shown in Fig.~\ref{fig:qca_maj}. In the depicted case,~a single \mbox{0-state} from input $a$ competes with two \mbox{1-states} coming from inputs $b$ and $c$. The output cell $f$ follows the majority of the input states and, thus, is forced to a \mbox{1-state} in this case. By locking the polarization of one of the three inputs to a constant state, one obtains a binary OR gate (constant 1-state) or an AND gate (constant 0-state), respectively.

%
%

In order to execute these and more complex logic operations, a dedicated \emph{clocking} is required which, starting with the initialization of the QCA cells, properly propagates among the cells and avoids metastable states~\cite{Liu2013design}. To this end, an external clock is employed which regulates the intercellular tunneling barriers within a QCA cell such that the cell can be polarized (i.\,e., tunneling is prevented) or not (i.\,e., electrons may tunnel between adjacent quantum dots within the cell). 
Typically, the clock consists of four phases: In the so-called \emph{relax} phase, the cell is depolarized and does not contain any information. During the following \emph{switch} phase, the interdot barriers are raised which forces the cell to polarize into one of the two antipodal states (according to the polarization of surrounding cells). In the following \emph{hold} phase, the cell keeps its polarization and may act as input for adjacent cells. During the final \emph{release} phase, the interdot barriers are lowered again thereby removing the previous polarization of the cell. 


In order to enable the propagation of information among cells, multiple phase-shifted versions of the clock signal are provided~\cite{Hennessy01}. 
Then, the data flow can be controlled by using appropriately shifted clock signals such that the cells which shall pass their data are in the  \emph{hold} phase at the same time when the cells that shall receive the data are in the \emph{switch} phase. Typically, multiple adjacent cells are grouped to \emph{clock zones} in which all cells use the same clock signal.

\begin{figure}
	\centering
	\includegraphics[scale=1]{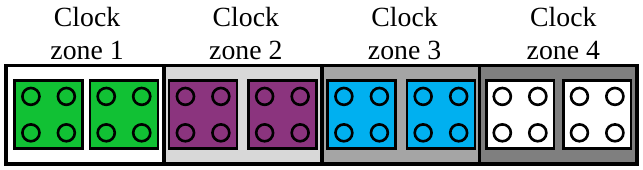}
	\caption{ QCA wire with cells in four clock zones }
	\label{fig:qca_clocks}
\end{figure}

\begin{example}
	Consider Fig.~\ref{fig:qca_clocks} showing a QCA wire consisting of eight cells. The cells are divided into four groups of pairs of adjacent cells which share the same clock signal. Each of these clock zones follows a different clock signal. More precisely, clock zone 2 will be in the \emph{switch} phase, when clock zone 1 is in the \emph{hold} phase. Thus, in this clock phase, cells in clock zone~2 polarize according to the polarization of the adjacent cells in clock zone 1. During the next clock phase, clock zone~2 changes to \emph{hold}, while clock zone 3 is in the \emph{switch} phase. Consequently, data is passed from zone 2 to 3 (and so on), similar to a pipeline structure. 
\end{example}

\subsection{Modeling the Energy Behavior of QCA Cells}
\label{sec:emodel}

Following the cyclic behavior of the clock signals, the energy behavior of QCA cells can be described in a qualitative way as follows. 
In the beginning of a clock cycle, the QCA cell is depolarized. Energy is taken from the clock as well as from neighboring cells in order to achieve the polarized state induced by the polarization of the neighboring cells. Most of this energy is restored to the clock as well as distributed to the neighboring cells until the cell becomes depolarized again at the end of the clock cycle. However, some portion of the energy dissipates to the environment.

In order to allow for a more precise, quantitative analysis of the energy behavior of QCA cells, one has to make use of the quantum-level modeling of QCA cell behavior discussed in several previous works, e.\,g. \cite{Timler02,Bhanja09,Sill18,Rahimi16}. 
In this model, the QCA cell behavior is determined by two \mbox{three-dimensional} energy vectors $\vec{\lambda}$ and $\vec{\Gamma}$. More precisely,  
the vector $\vec{\lambda}=(\lambda_x, \lambda_y, \lambda_z)$ denotes the so-called \emph{coherence vector} and represents the cell's current state (where $\lambda_z$ can be identified with its polarization). In a similar fashion, the energy vector $\vec{\Gamma}=\frac{1}{\hbar}\left[ -2\gamma,0,\Phi \right]$, with $\hbar$ denoting the reduced Planck constant, is related to the cell's \emph{steady-state}, i.\,e. a virtual state that characterizes the future behavior of the cell and depends on the current tunneling behavior ($\gamma$) as well as the Coulomb force  that is induced by neighboring cells~($\Phi$).

\medskip 
Using this notation, the instantaneous power $P$ of a QCA cell is described by

\begin{equation}
\label{eq:Power}
P = \frac{d}{dt}\mathbb{E}(t) = \frac{d}{dt} \left( \frac{\hbar}{2}\vec{\Gamma}(t) \cdot \vec{\lambda}(t) \right),
\end{equation}
where $\mathbb{E}(t)$ denotes the current energy of the cell at time $t$ and is essentially given as the scalar product of the two energy vectors at that point in time. 
%
%
%
Consequently, the total energy dissipation of a QCA cell ($E_{total}$) during a complete clock cycle with period $T_{clk}$  is given as
\begin{equation}
\label{eq:Eall}
\medmuskip=0mu
\thinmuskip=0mu
\thickmuskip=-1mu
E_{total} = \int_{t_0}^{t_0+T_{clk}} P dt' =  \frac{\hbar}{2} \int_{t_0}^{t_0+T_{clk}} \left( \frac{d}{dt}\vec{\Gamma} \cdot \vec{\lambda} + \frac{d}{dt}\vec{\lambda} \cdot \vec{\Gamma}  \right) dt'.
\end{equation}

The first summand in the integrand of Eq.~(\ref{eq:Eall}) is the scalar product of the derivate of the energy vector of the cell ($\vec{\Gamma}$) and the coherence vector. This term refers to the energy transfer with the clock ($E_{clk}$) as well as neighboring cells ($E_{IO}$) during a clock cycle~\cite{Bhanja09,Timler02,Sill18}, i.\,e.
\begin{equation}
E_{clk} + E_{IO} = \frac{\hbar}{2} \int_{t_0}^{t_0+T_{clk}} \left( \frac{d}{dt}\vec{\Gamma} \cdot \vec{\lambda} \right)dt'
\end{equation}  
with
\begin{equation}
\label{eq:E_clkc}
E_{clk} = \frac{1}{2} \int_{t_0}^{t_0+T_{clk}} \left( \frac{d}{dt} (-2\gamma) \cdot \lambda_x \right)dt'
\end{equation}
and   
\begin{equation}
\label{eq:E_ioc}
E_{IO} = \frac{1}{2} \int_{t_0}^{t_0+T_{clk}} \left( \frac{d}{dt} \Phi \cdot \lambda_z \right)dt'.
\end{equation}

The second summand in the integrand of Eq.~(\ref{eq:Eall}) is the product of the derivate of the coherence vector and the energy vector and captures the energy transfer $E_{env}$ with the environment during a clock cycle. $E_{env}$ is the actually dissipated energy of a QCA cell during a clock cycle and can be determined as 
\begin{equation}
\label{eq:E_envc}
\begin{split}
E_{env} = -\frac{\hbar}{2\tau} \int_{t_0}^{t_0+T_{clk}}    \left[ \left( \vec{\Gamma} \cdot \vec{\lambda} + |\vec{\Gamma}| \tanh \eta_{th}  \right) \right] dt', \\
\end{split}
\end{equation}
where $\tau$ denotes a technology-dependent relaxation time parameter and  \mbox{$\eta_{th} = \hbar \lvert\vec{\Gamma}\rvert\cdot(2k_BT)^{-1}$} refers to the thermal ratio~\cite{Timler02,Bhanja09,Sill18}.




One should note that the proposed model assumes an ideal clock signal that consumes no energy. As discussed in~\cite{Lent06}, there occurs indeed some heat dissipation in the clocking circuit due to wire resistances and the clock generation circuitry. However, this is not a fundamental limitation and can be minimized by reducing the residual resistances~\cite{Lent06,Jeanniot16}.

\subsection{Simulating the Energy Dissipation of QCA}
\label{sec:QD-E}

Currently, there are two tools available that allow for the simulation of the energy dissipation in QCA circuits---\textit{QCADesigner-E}~\cite{Sill18} and QCAPro~\cite{Bhanja09}. However, the latter is not able to estimate exact values as it assumes an ideal clock slope, and thus, only allows for determining an upper-limit for the energy dissipation. In contrast, the open-source simulation tool \textit{QCADesigner-E}\footnote{The tool has been made publicly available as open-source at~\url{https://github.com/FSillT/QCADesigner-E}} permits an precise estimation.
It is based on the widely applied \textit{QCADesigner}~\cite{Walus06} and computes the energy terms $E_{clk}$, $E_{IO}$, and $E_{env}$ from Eq.~\eqref{eq:E_clkc}--\eqref{eq:E_envc} by interpolating the respective integrands.
To this end, the required numerical values of the energy vectors $\vec{\lambda}$ and $\vec{\Gamma}$ are obtained by employing the \emph{Coherence Vector Simulation Engine} (CVSE) that is included in the original \textit{QCADesigner} tool. This engine implements the state-of-the-art quantum-level modeling of QCA cell behavior discussed in~\cite{Timler02,Bhanja09,Sill18} and determines the evolution of $\vec{\lambda}$ and $\vec{\Gamma}$ by solving the corresponding differential equations using an iterative, fixed timestep approach. The time interval of each iteration step ($T_{step}$) as well as many other technology and simulation parameters (shown in Table~\ref{tab:qcad_params})
can be adjusted in order to adapt to a specific physical realization or to increase the precision of the simulation. In fact, as the simulation error that may occur due to inadequate iteration step lengths or rounding is permanently tracked, a simulation run can be repeated with higher resolution if the error grows too large.
However, as demonstrated and validated in~\cite{Sill18}, using the standard simulation parameters of the tool (listed in Table~\ref{tab:qcad_params}) one can in general expect high quality results with simulation errors below~5\%.

\setlength{\tabcolsep}{4pt}

\begin{table}[t]
	\centering
	\caption{Adjustable technology and simulation parameters in the tool \textit{QCADesigner-E}}
	\label{tab:qcad_params}
	{\renewcommand{\arraystretch}{01}
		\subfloat[Technology parameters\label{tab:qcad_params_std}]{
			\begin{tabular}{|c|p{4cm}|c|}
				\hline
				\textbf{Parameter} & \textbf{Description} & \textbf{Standard Value} \\ \hline
				QD size 	& Size of a quantum dot & 5~nm \\
				Cell area 	& Dimensions of each cell & 18~nm x 18~nm\\
				Cell distance & Distance between two cells & 20~nm\\
				Layer distance & Distance between QCA layers in case of multi-layer crossing~\cite{Bajec12} & 11.5~nm \\
				$\tau$ 		& Relaxation time & 1E-15~s \\
				$\gamma_H$	& Max. saturation energy of clock signal & 9.8E-22~J\\
				$\gamma_L$	& Min. saturation energy of clock signal & 3.8E-23~J\\
				$\epsilon_r$& Relative permittivity of material for QCA system & 12.9$^*$ \\ 
				Temp 		& Operating temperature & 1~K \\ 
				$r_{\textrm{effect}}$& Maximum distance between cells whose interaction is considered & 80~nm$^\dagger$ \\ \hline
				\multicolumn{3}{l}{\scriptsize
					\begin{tabular}[t]{l@{\hskip 2pt}l} 
						$^*$ &Relative permittivity of GaAs and AlGaAs \\ 
						$^\dagger$ &Interaction effects between two cells decays inversely with the fifth \\ &power of its distance
				\end{tabular}}
		\end{tabular}}
		
		\subfloat[Simulation parameters]{ 
			\begin{tabular}{|c|p{4cm}|c|}
				\hline
				\textbf{Parameter} & \textbf{Description} & \textbf{Standard Value} \\ \hline
				$T_{\gamma}$& Period of the clock signal & 10E-12~s\\
				$\gamma_{slope}$ & Rise and fall time of the clock signal slopes & 1E-10 s\\			
				$\gamma_{shape}$ & Shape of clock signal slopes [RAMP$/$GAUSSIAN] & GAUSSIAN \\
				$T_{in}$& Period of the input signals & 10E-12~s\\
				$T_{sim}$	& Total simulation time &	80E-12~s\\
				$T_{step}$	& Time interval of each iteration step & 1E-17~s \\
				\hline	
	\end{tabular}}}
	
\end{table}

\section{Energy Dissipation in QCA Building Blocks}\label{sec:bb}

Using the simulator reviewed above as basis, this section describes the obtained energy dissipation for common QCA building blocks. 
The main subject of our study is to validate whether QCA designs that are expected to be reversible indeed allow for operating below the $(k_BT \ln 2)$ energy limit or not. 
To this end, we investigate a single QCA wire as well as an elementary logic operation (namely an OR) as proper representatives. While the wire inherently is reversible at the logic level, the OR operation obviously is not. However, findings summarized in~\cite{Lent06} suggested a corresponding QCA design which is expected to achieve reversibility, both logical \emph{and} physical. 
This design is, of course, also considered within this study. 
Please note that the issue of cooling will be ignored in the following, having in mind that this work is a theoretical analysis exploring whether the actual concept of QCA enables near zero-energy dissipation computing. Extrapolating any costs for cooling costs could easily yield to misleading conclusions.

\subsection{Energy Dissipation in QCA Wires}\label{sec:wire_qca}

As discussed in Section~\ref{sec:emodel}, there is an energy transfer amongst QCA cells~(denoted by $E_{IO}$), from the clock to the QCA cells~(denoted by $E_{clk}$) and from the QCA cells to the environment~(denoted by $E_{env}$).  
Timler and Lent discussed that energy coming from the clock is used to restore the signal strength of internal signals and compensate energy dissipated to the environment~\cite{Timler02,Timler03}. This shall be detailed in the following example.

\begin{figure}
	\centering
	\includegraphics[scale=0.8]{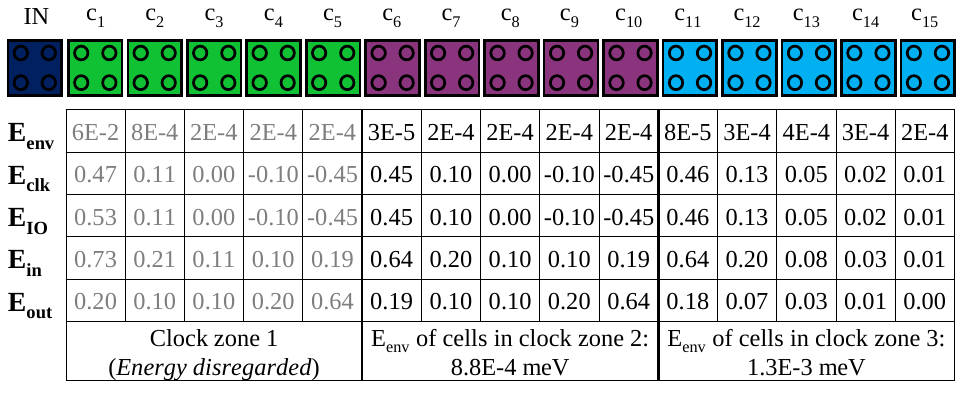}
	\caption{Energy transfer in QCA wire. All values in meV. \label{fig:energy_wire}}
\end{figure}

\begin{example}
	\label{ex:energy_wire}
	The top of Fig.~\ref{fig:energy_wire} shows a QCA wire composed of cells in three clock zones. The corresponding energy transfer between cells, clock, and environment is summarized in the bottom of Fig.~\ref{fig:energy_wire}. To obtain these numbers, the wire has been simulated with the \mbox{QCADesigner-E} tool using its standard parameters (listed in Table~\ref{tab:qcad_params}).
	
	The symbols $E_{in}$ and $E_{out}$ represent the energy entering~($E_{in}$) and leaving~($E_{out}$) the QCA cell. In case of the depicted wire, the value for $E_{IO}$ follows from the difference between $E_{out}$ and $E_{in}$. Further, positive values mean that energy has been transferred to the environment ($E_{env}$), to the clock ($E_{clk}$), and to the cell ($E_{IO}$), while negative values mean the opposite. In order to neglect any effect due to the ideal signal source (marked as \emph{IN}) whose behavior is to some extent artificial, the energy of  QCA cells in zone~1 (highlighted in green) shall be ignored. Nevertheless, the behavior of all cells has been fully simulated. 
		
	As can be seen, energy is transferred from the clock to the QCA cells in order to compensate energy transferred to the next cell. This can been seen e.\,g.~at cell $c_{10}$, which receives 0.19~meV from its left neighbors and passes 0.64~meV to its right neighbors located in the next clock zone. The resulting difference of -0.45~meV is then compensated by the clock signal.
	Similarly, all energy not transferred to neighboring cells is sent back to the clock. 
	For example, cell $c_{11}$ receives 0.64~meV from its left neighbors, but passes only 0.18~meV to its right neighbors. The difference of 0.46~meV is returned to the clock.	
	As expected, the amount of energy $E_{env}$ that is transferred to the environment is negligible and clearly stays below the $(k_BT \ln 2)$ energy limit of 0.06~meV, a mandatory requirement for designing circuits that are able to operate with near zero-energy dissipation.
	
\end{example}  

As can be seen from Eq.~\eqref{eq:E_envc} and as already discussed before in~\cite{Bhanja09} and~\cite{Timler02}, the energy dissipation of a QCA cell mainly results from the difference between its current state (represented by the coherence vector $\vec{\lambda}$) and its steady-state (represented by the energy vector $\vec{\Gamma}$).
As outlined in Section~\ref{sec:emodel}, this difference results from changing polarizations of neighboring cells, leading to changing Coulomb forces, as well as from the varying tunneling barrier, both directly related to the clock signal.
As a consequence, the inclination of the slope of the external clock signal has a considerable impact on the energy transfer to the environment, i.\,e. the total energy dissipation of the QCA cell. In other words: the slower the clock signal changes, the easier it is for the cell to follow the (accordingly changed) steady-state and the less energy is dissipated.
This behavior has already been discussed by Bhanja et al. in~\cite{Bhanja09} and is illustrated by the following example.

\begin{figure}
	\centering
	\includegraphics[scale=0.75]{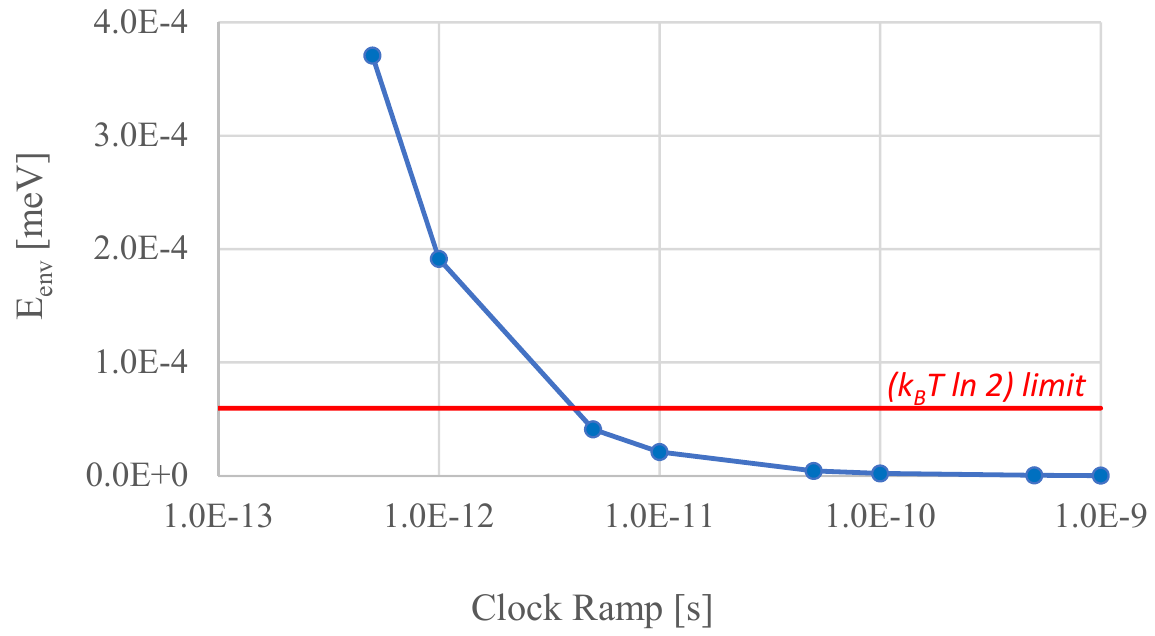}
	\caption{Energy dissipation $E_{env}$ of a QCA wire in relation to the slope of the external clock signal\label{fig:energy_slope}}
\end{figure}

\begin{example}
Figure~\ref{fig:energy_slope} depicts the relation between the slope of the external clock signal~$\gamma_{slope}$ and the energy transfer to the environment $E_{env}$ of the wire shown in Fig.~\ref{fig:energy_wire}. The numbers were generated using the \textit{QCADesigner-E} tool and the standard parameters (listed in Table~\ref{tab:qcad_params_std}).
The results indicate an exponential relation between the clock slope~$\gamma_{slope}$ and~$E_{env}$. 
It follows further that clock slopes longer than 10~ps lead to operations with energy dissipation below the $(k_BT \ln 2)$ limit.
\end{example}

Overall, the conducted case studies confirm that the QCA wire indeed can be operated below the $(k_BT \ln 2)$ limit, a fundamental requirement for building circuits that are logically and physically reversible.

\subsection{Energy Dissipation in QCA Gates}\label{sec:bit_erasure}

Next, we consider the realization of QCA gates---with an OR operation as representative. The OR is a \mbox{non-reversible} logic operation which may cause information-loss and, hence, yields an energy dissipation above Landauer's limit.
Lent discussed that this process can be turned reversible for a specific cell, if first a copy of the information (bit) is made~\cite{Lent06}.
When the cell is then actually losing its information, that copy would act as a \emph{Demon}---a hypothetical model introduced by Maxwell in 1875~\cite{Lent06}---and, thus, would allow to return the cell to a null state without thermodynamical energy loss.

Timler et al. discussed in \cite{Timler03} that, in case of QCA, the information erasure happens during the \emph{release} phase, i.\,e.~when the interdot barriers are lowered and the polarization of the QCA cell is removed (see also Section~\ref{sec:basics_QCA}). Further, they observed that a neighboring QCA cell can act as the \emph{Demon}, i.\,e.~as a holder of the copy of the bit which is stored as polarization. 

The results presented in \cite{Timler03} and in the above Example~\ref{ex:energy_wire} confirm this observation. One can observe that during the \emph{release phase} all cells within the same clock zone lose their polarization, i.\,e.~the stored information. However, for the QCA wire there is no energy transfer to the environment, i.\,e.~this erasure process happens in a reversible fashion.

In case of a standard QCA OR gate it is not possible to assure for all possible input patterns that such a copy always exists, i.\,e. that neighboring cells possess the same information. The following example shall detail the consequences.

\begin{figure}[t]
	\centering
	\subfloat[Standard QCA OR \label{fig:or_standard}]{\hspace{3mm}\includegraphics[scale=0.8]{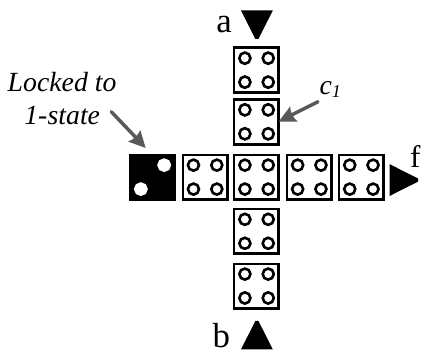}\hspace{-1mm}} \hfil
	\subfloat[Reversible QCA~OR\label{fig:or_reversible}]{\includegraphics[scale=0.8]{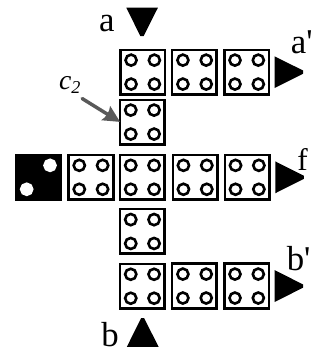}\hspace{5mm}}
	
	\subfloat[Related clock, $P_{env}$ and polarization of cell $c_1$ in standard OR and cell $c_2$ in reversible OR. The $(k_BT \ln 2)$ limit is at 0.06~meV. \label{fig:or_curves}]{\includegraphics[scale=0.7]{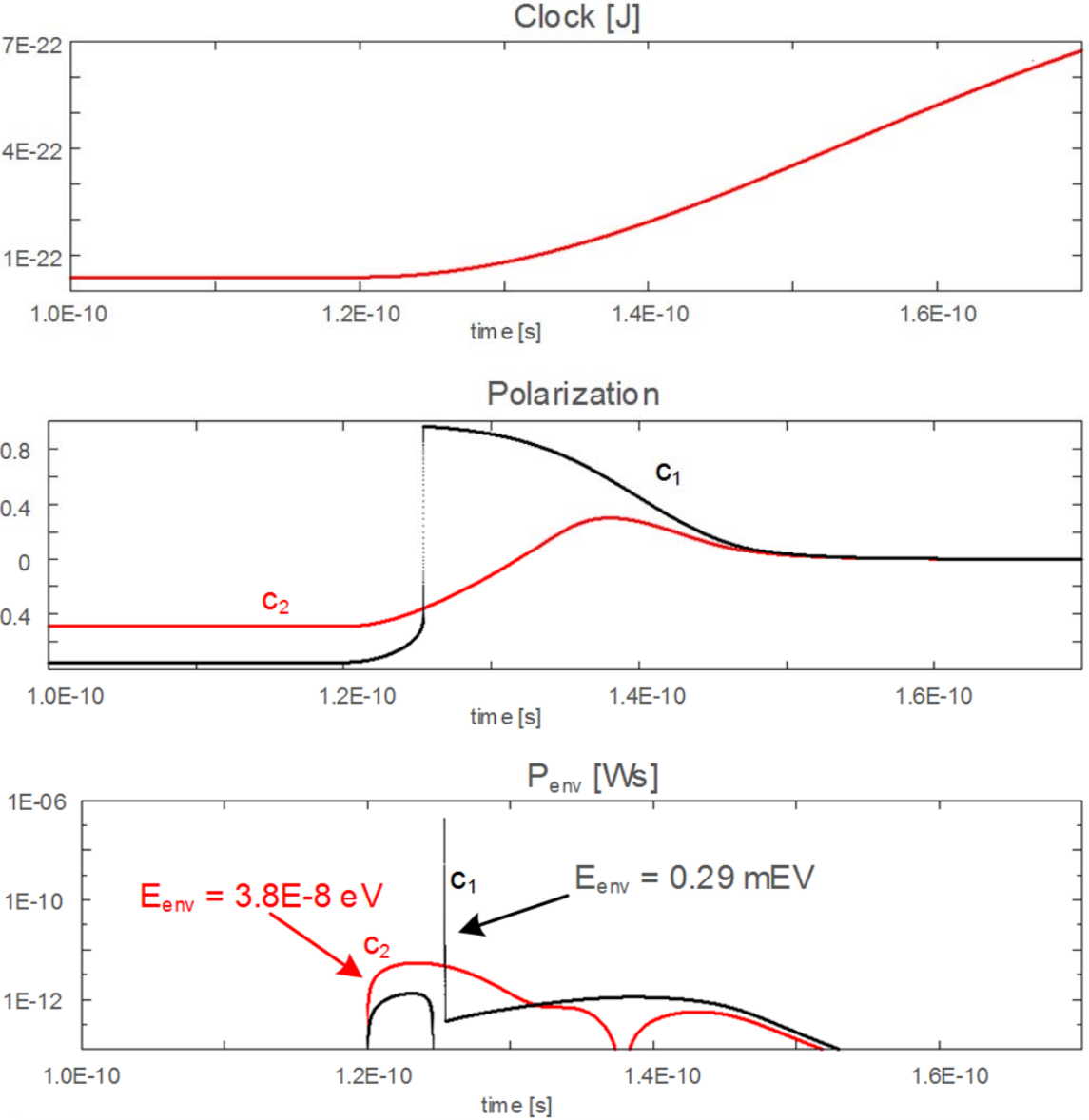}} 
	\caption{Bit erasure in standard and reversible QCA OR}\label{fig:erasure_or}
\end{figure}

\begin{example}\label{ex:sim_or}
	Consider the standard QCA OR gate depicted in Fig.~\ref{fig:or_standard}, which has been simulated with the \mbox{QCADesigner-E} tool using the standard parameters listed in~
	Table~\ref{tab:qcad_params_std} for the case $a=0$ and $b=1$.
	Fig.~\ref{fig:or_curves} shows the integrand $P_{env}$ of the energy dissipated to the environment (see also Eq.~\ref{eq:E_envc}), the polarization and the related clock signal for the cell $c_1$. These results indicate that, during \emph{release phase}, the polarization of cell $c_1$ changes abruptly---leading to a high energy transfer to the environment and, consequently, to an energy dissipation which is above the $(k_BT \ln 2)$ limit. 
\end{example}

However, following the concept of demon cells discussed above, the high energy dissipation can be prevented by using a modified OR that echoes both inputs to the output of the cell.
This approach has originally been proposed in~\cite{Lent06} and is expected to result in a reversible erasure with an energy dissipation below the $(k_BT \ln 2)$ limit.
However, the actual energy behavior of the proposed design has not been evaluated with precise simulations yet.

\begin{example}
	Fig.~\ref{fig:or_reversible} shows the modified OR gate as proposed in~\cite{Lent06}. Simulating this gate with the \textit{QCADesigner-E} tool using the same settings as reported in Example~\ref{ex:sim_or} leads to results as summarized in Fig.~\ref{fig:or_curves}. Here, it can be seen that, during the \emph{release phase}, no abrupt change of the polarization of cell $c_2$ occurs. Consequently, the energy transfer to the environment is considerably lower. 
	In fact, an energy dissipation can be observed which is always below the $(k_BT \ln 2)$ energy limit, i.\,e. the computation of the reversible QCA OR gate is indeed logically and physically reversible.\footnote{Note that, in order to achieve reversibility at the logic level, the primary inputs are echoed to the primary outputs of the design. This increases the number of primary outputs to three. But still, the gate does not seem to be reversible at the logic level (there are now more outputs than inputs) and, thus, Landauer's principle seems to contradict the physical reversibility determined by our simulations. However, this contradiction is resolved if the locked cell is interpreted as a third---but constant---primary input~\cite{zulehner17}.}
\end{example}

For the first time, these results confirm the hypotheses of~\cite{Lent06,Huang06} with precise simulation results and, by this, validate the general capability of QCA to implement logic operations reversibly at both, the logic and the physical level,
i.\,e. a logically reversible operation can be conducted with an energy dissipation below the $(k_BT \ln 2)$ limit. Next, we evaluate whether physical reversibility can also be achieved for larger QCA designs.


\section{Energy Dissipation in Larger QCA Designs}\label{sec:results}

The investigations summarized above showed that QCA building blocks indeed allow for (1)~the realization of established functional building blocks in a reversible fashion (including maintaining reversibility at the physical level) and, hence, (2)~an energy dissipation which is below the $(k_BT \ln 2)$ energy limit.  Now, these findings are employed to investigate whether similar results can also be observed for larger QCA designs as well.
In this section, we first summarize the case studies which have been conducted to evaluate this. Afterwards, the obtained results are presented and discussed.

\subsection{Conducted Case Studies}

\begin{figure}
	\centering
	\subfloat[Proposed reversible Majority Gate \label{fig:MAJ_rev}]{\hspace{10mm}\includegraphics[scale=1]{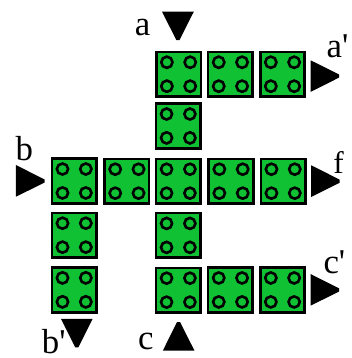}\hspace{10mm}} \hfil
	\subfloat[Reversible Half-Adder ($a_{cp}$ and $b_{cp}$ indicate copies of the inputs, $g_1$ and $g_2$ are so-called 
	\emph{garbage outputs})\label{fig:HA_rev}]{\includegraphics[scale=.5]{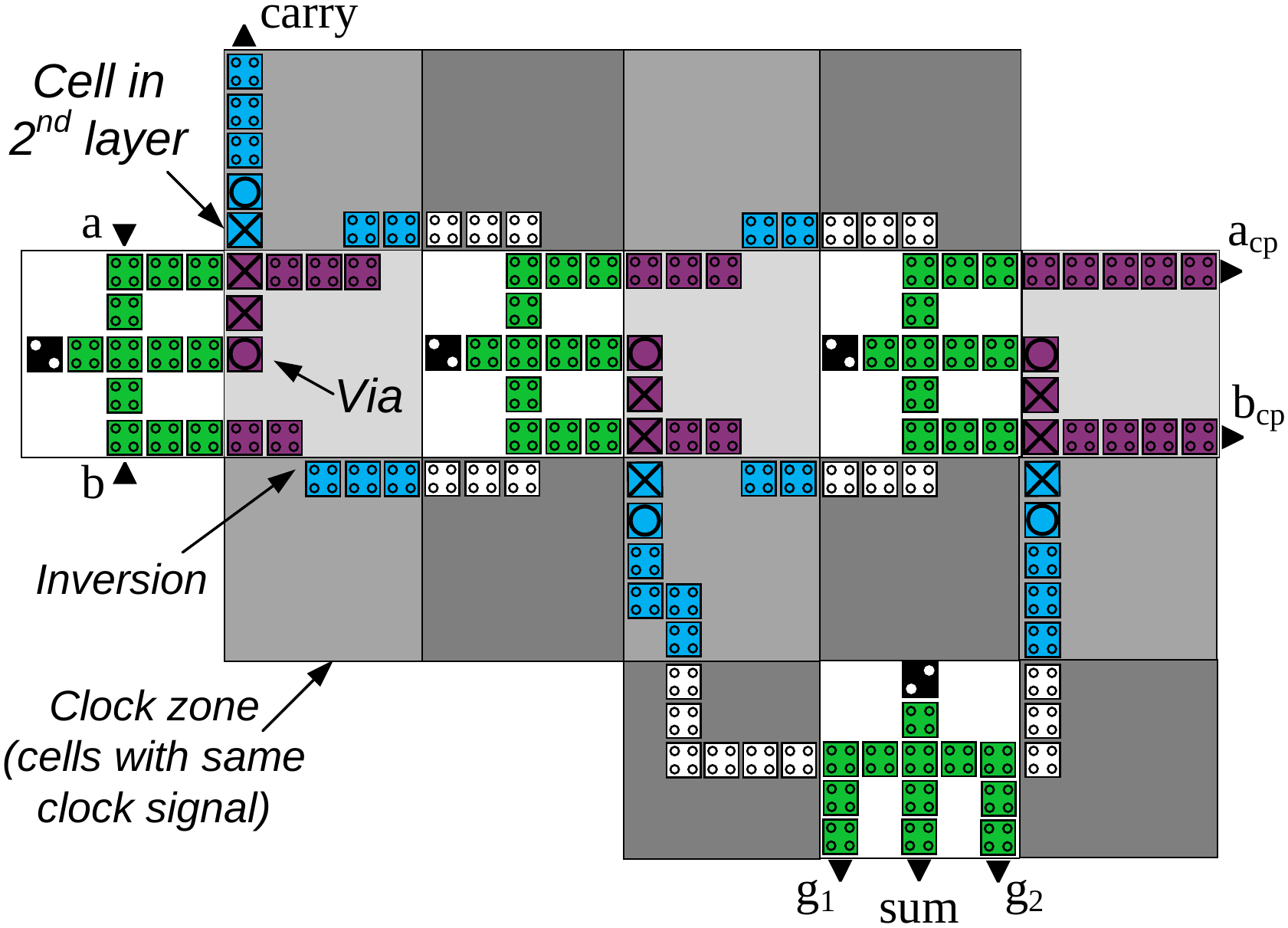}} \hfil
	\caption{Considered logically and physically reversible QCA Designs}\label{fig:qca}
\end{figure}

As larger QCA designs, we considered the realization of QCA implementations from the following three categories:
\begin{itemize}
\item \emph{Standard QCA Designs}, i.\,e.~QCA implementations which have been commonly used for QCA designs. More precisely, the standard OR gate, the standard AND gate and the standard Majority gate are considered as representatives for this category.

\item \emph{Logically (but not physically) Reversible QCA Designs}, i.\,e.~implementations as proposed in related works
which implement reversible logic functions, but, as discussed in Section~\ref{sec:intro}, have either no or only a ``conventional'' (i.\,e.~\mbox{non-reversible}) physical implementation. More precisely, an implementation of the Feynman gate~\cite{Chaves15,Singh17}, a logically reversible \mbox{T-FlipFlop} (\mbox{T-FF})~\cite{Mukhopadhyay15}, a testable adder implemented in reversible logic (T-adder)~\cite{Sen17}, as well as a circuit based on multiplexers and XOR gates implementing a random reversible function (RevFunc)~\cite{Chabi17} are considered as representatives for this category. 

\item \emph{Logically and Physically Reversible QCA  Designs}, i.\,e.~implementations based on the schemes discussed in~\cite{Lent06} and reviewed in the previous section which are reversible 
and, at least from a theoretical point of view, can be operated below the $(k_BT \ln 2)$ energy limit. 
More precisely, the OR gate implementation from Fig.~\ref{fig:or_reversible}, its counterpart the reversible AND gate\footnote{The reversible AND gate is nearly identical to the reversible OR gate depicted in Fig.~\ref{fig:or_reversible}, except for the constant input that is instead locked to the 0-state.} as well as a Majority gate implementation\footnote{The difference in the number of inputs (3) and outputs (4) does not change the fact that this gate can be understand as reversible circuits. This follows from Landauer’s observation that the fundamental energy dissipation is defined by the change of entropy, i.\,e. the difference between initial entropy $S_i$ (before operation) and final entropy $S_f$ (after operation)~\cite{DeBenedictis16}. In case of the proposed gate, both entropies $S_i$ and $S_f$ are identically, as each possible output state is related to one and only one input state, while all remaining output states have a probability of zero.}
proposed in Fig.~\ref{fig:MAJ_rev} are considered as representatives for this category. 


\end{itemize}

Further, we propose an implementation of a logically and physically reversible Half-Adder (see Fig.~\ref{fig:HA_rev}) that, for the first time, utilizes exclusively logically \emph{and} physically reversible circuit elements, e.\,g. as introduced by Lent et al.~\cite{Lent06} and in this work. The wire-crossings (indicated by QCA cells containing crosses in Fig.~\ref{fig:HA_rev}) are realized via the multilayer approach discussed in~\cite{Bajec12}\footnote{Despite the fact that multilayer design is not trivial, the concept has been explored in several previous works with very favorable conclusion regarding its feasibility~\cite{Bajec12,Tougaw99}}.
The design of this circuit is based on a tile-based design approach as proposed in~\cite{Huang05}, which facilitates an automatic implementation as, for example, shown by~\cite{Walter18,Wille19,Walter19b}.

All evaluations have been executed with the same method and settings as done before for the case studies in Section~\ref{sec:bb}, i.\,e.~with the tool \textit{QCADesigner-E}~\cite{Sill18} reviewed in Section~\ref{sec:emodel} using its standard parameters listed in Table~\ref{tab:qcad_params_std}.
The clock slope was set to $\gamma_{slope}=100~ps$, following from the observations in Section~\ref{sec:wire_qca} and Fig.~\ref{fig:energy_wire}. The time interval $T_{step}$ of each iteration step 
was set to 0.01~fs---leading to simulation errors of~$\epsilon_{env}\leq1~\%$. 
As proposed in \cite{Sill18}, we cascaded the artificial input signals by placing a few buffer cells between the stimulated inputs and the actual inputs of the considered design (in order to allow for a realistic energy analysis). The same is done for the outputs, which are as well connected to a few buffer cells.

\begin{table*}[t]
	\centering
	\caption{Energy dissipation in larger QCA designs}
	\resizebox{\columnwidth}{!}{
	\label{tab:results}
	\begin{tabular}{|l||c|c|c|c|c|c|c|c|}
		\hline
		\multicolumn{1}{|l||}{\emph{Circuits}}  & \multicolumn{8}{c|}{\emph{Energy Dissipation [meV] for an Input Signal Combination}}                                                                                         \\ 
		& 000            & 001            & 010            & 011            & 100            & 101            & 110            & 111            \\ \hline\hline	
		\multicolumn{9}{|l|}{\emph{Standard QCA Designs}}   \\\hline                         	
		Inverter                   & \textbf{0.001}          &  \textbf{0.001}            &                &                &                &                &                &                \\ \hline
		OR (standard)              & 0.082 & 0.709          & 0.712          & \textbf{0.001} &                &                &                &                \\ 
		AND (standard)              & \textbf{0.001} & 0.711          & 0.709          & 0.081 &                &                &                &                \\ 
		Majority (standard)        & \textbf{0.001} & 0.709          & 0.714          & 0.711          & 0.709          & 0.714          & 0.711          & \textbf{0.001} \\ \hline \hline
		
		\multicolumn{9}{|l|}{\emph{Logically (but not physically) Reversible QCA Designs}}   \\\hline
		Feynman \cite{Chaves15}             & 0.817          & 1.645          & 1.645          & 0.867          &                &                &                &                \\ 
		Feynman \cite{Singh17}            & 0.393          & 0.064 & 0.839          & 1.376          &                &                &                &                \\ 
		T-FF \cite{Mukhopadhyay15}              & 1.053          & 0.397          & 2.072          & 2.200          & 2.993          & 1.424          & 1.050          & 1.138          \\ 
		RevFunc \cite{Chabi17} & 1.190          & 0.742          & 0.753          & 0.353          & 1.537          & 0.130          & 0.684          & 0.074 \\ 
		T-adder \cite{Sen17}              & 1.565          & 1.565          & 2.458          & 2.493          & 1.315          & 1.315          & 2.493          & 2.458          \\ 
		\hline \hline
		
		\multicolumn{9}{|l|}{\emph{Physically Reversible QCA  Designs}}   \\\hline
		OR (reversible, Fig.~\ref{fig:or_reversible})            & \textbf{0.002} & \textbf{0.003} & \textbf{0.002} & \textbf{0.002} &                &                &                &                \\
		AND (reversible)            & \textbf{0.002} & \textbf{0.002} & \textbf{0.003} & \textbf{0.002} &                &                &                &                \\  
		Majority (reversible, Fig.~\ref{fig:MAJ_rev})      & \textbf{0.003} & \textbf{0.003} & \textbf{0.003} & \textbf{0.003} & \textbf{0.003} & \textbf{0.003} & \textbf{0.003} & \textbf{0.003} \\ 
		Half-Adder (Fig.~\ref{fig:HA_rev})                & \textbf{0.022} & \textbf{0.025} & \textbf{0.029} & \textbf{0.022} &
		&                &                &                \\  \hline
	\end{tabular}
	}

	{\scriptsize	
		\vspace{2mm}
		$(k_BT \ln 2)$ energy limit is at 0.06~meV (values marked bold are below this limit)\\
		Simulations executed with Gaussian-shaped clock slope of $\gamma_{slope}=100~ps$ and Temp=1~K
	
	}

\end{table*}

\subsection{Results and Discussions}
Table~\ref{tab:results}  lists the obtained energy dissipation for each input signal combination (in meV). Values marked in bold indicate an energy dissipation below the $(k_BT \ln 2)$ energy limit (which is at 0.06~meV).
These results allow for the following conclusions:

\begin{itemize}
\item Already the standard and the logically reversible QCA designs allow, in very few cases, for an energy dissipation below the $(k_BT \ln 2)$ barrier.
This, however, is only the case when input assignments are employed which do not yield to an information loss.
This is e.\,g.~the case for the OR implementation with the input assignment '11' (see e.\,g. Fig.~\ref{fig:or_standard}).

\item In all other cases, we could confirm that logical reversibility is not sufficient to operate below the $(k_BT \ln 2)$ energy barrier. As already discussed above, this is because these realizations are based on \mbox{non-reversible} structures, e.\,g. standard AND, OR, Majority gates, and, hence, employ a ``conventional'' (i.\,e.~\mbox{non-reversible}) physical realization. As a consequence, only for few selected input assignments which, similar to '11' for the OR gate, do not cause any information loss, an energy dissipation below the  $(k_BT \ln 2)$ limit is possible.
For all other cases, the limit can never be broken which is also confirmed by the reported results.

\item In contrast, we are able to confirm that logically and physically reversible implementations indeed allow for an energy dissipation below $(k_BT \ln 2)$ per operation, and thus, near zero-energy computing. In fact, for all corresponding implementations, energy dissipations which are clearly below the $(k_BT \ln 2)$ barrier are reported. This is also the case for the proposed reversible Majority gate and the Half-Adder circuit. By this, for the first time we could show with precise simulations that it is indeed possible to implement QCA designs that satisfy the requirements for physically reversible circuits and, hence, allow for near zero-energy computing.
\item Furthermore, as we could show that it is possible to implement all fundamental Boolean operations (i.\,e. AND, OR, and inversion) in a logically and physically reversible manner, one can conclude that all Boolean logic functions can be designed as logically and physically reversible QCA designs. 
\end{itemize}

\section{Conclusions}\label{sec:concl}
In this work, we validated that it is possible to design QCA circuits which are logically and physically reversible, and thus, allow for near zero-energy computing.
To this end, we conducted dedicated case studies on corresponding QCA designs for primitive building blocks such as wires or functional gates (like OR) as well as larger QCA designs. Furthermore, we applied reversible building blocks and a common design methodology for QCA in order to develop the first logically and physically reversible QCA adder circuit. The obtained results  confirm the general capability of QCA designs to compute with near zero-energy dissipation.
This constitutes an essential result which further motivates the consideration of QCA as an alternative to overcome physical limitations of conventional computation technologies. Future work is  defined by confirming these findings on physically built QCA designs---the results obtained in this work deliver a strong motivation towards this path.

	
\bibliographystyle{IEEEtran}
\bibliography{ms}

\begin{thebibliography}{10}
\providecommand{\url}[1]{#1}
\csname url@samestyle\endcsname
\providecommand{\newblock}{\relax}
\providecommand{\bibinfo}[2]{#2}
\providecommand{\BIBentrySTDinterwordspacing}{\spaceskip=0pt\relax}
\providecommand{\BIBentryALTinterwordstretchfactor}{4}
\providecommand{\BIBentryALTinterwordspacing}{\spaceskip=\fontdimen2\font plus
\BIBentryALTinterwordstretchfactor\fontdimen3\font minus
  \fontdimen4\font\relax}
\providecommand{\BIBforeignlanguage}[2]{{%
\expandafter\ifx\csname l@#1\endcsname\relax
\typeout{** WARNING: IEEEtran.bst: No hyphenation pattern has been}%
\typeout{** loaded for the language `#1'. Using the pattern for}%
\typeout{** the default language instead.}%
\else
\language=\csname l@#1\endcsname
\fi
#2}}
\providecommand{\BIBdecl}{\relax}
\BIBdecl

\bibitem{Landauer61}
R.~Landauer, ``Irreversibility and heat generation in the computing process,''
  \emph{IJRD}, vol.~5, pp. 183--191, 1961.

\bibitem{Berut12}
A.~Berut \emph{et~al.}, ``{Experimental verification of Landauer's principle
  linking information and thermodynamics},'' \emph{Nature}, vol. 483, no. 7388,
  pp. 187--189, Mar. 2012.

\bibitem{Hong16}
J.~Hong, B.~Lambson, S.~Dhuey, and J.~Bokor, ``Experimental test of
  landauer{\textquoteright}s principle in single-bit operations on nanomagnetic
  memory bits,'' \emph{Science Advances}, vol.~2, no.~3, 2016.

\bibitem{Orlov12}
A.~O. Orlov, C.~S. Lent, C.~C. Thorpe, G.~P. Boechler, and G.~L. Snider,
  ``Experimental test of landauer's principle at the sub-kbt level,'' vol.~51,
  no.~6S, p. 06FE10, 2012.

\bibitem{Neri16}
I.~Neri and M.~{Lopez-Suarez}, ``Heat production and error probability relation
  in landauer reset at effective temperature,'' in \emph{Nature - Scientific
  reports}, 2016.

\bibitem{Gershenfeld:1996}
N.~Gershenfeld, ``{Signal entropy and the thermodynamics of computation},''
  \emph{ISL}, vol.~35, no. 3--4, pp. 577--586, 1996.

\bibitem{Bennett73}
C.~H. Bennett, ``Logical reversibility of computation,'' \emph{IBM J. Res.
  Dev.}, vol.~17, no.~6, pp. 525--532, Nov. 1973.

\bibitem{Zulehner19}
A.~Zulehner, M.~P. Frank, and R.~Wille, ``Design automation for adiabatic
  circuits,'' 2019, pp. 669--674.

\bibitem{ye1996energy}
Y.~Ye and K.~Roy, ``Energy recovery circuits using reversible and partially
  reversible logic,'' \emph{IEEE TCS}, vol.~43, no.~9, pp. 769--778, 1996.

\bibitem{zulehner2017one}
A.~{Zulehner} and R.~{Wille}, ``One-pass design of reversible circuits:
  Combining embedding and synthesis for reversible logic,'' vol.~37, no.~5, pp.
  996--1008, May 2018.

\bibitem{Chaves19}
J.~F. {Chaves}, M.~A. {Ribeiro}, F.~{Sill Torres}, and O.~P.~V. {Neto},
  ``Designing partially reversible field-coupled nanocomputing circuits,''
  \emph{IEEE Transactions on Nanotechnology}, vol.~18, pp. 589--597, 2019.

\bibitem{Frank17}
M.~P. Frank, ``Throwing computing into reverse: The future of computing depends
  on making it reversible,'' vol.~54, no.~9, pp. 32--37, September 2017.

\bibitem{DeBenedictis16}
E.~P. {DeBenedictis}, M.~P. {Frank}, N.~{Ganesh}, and N.~G. {Anderson}, ``A
  path toward ultra-low-energy computing,'' in \emph{2016 IEEE International
  Conference on Rebooting Computing (ICRC)}, Oct 2016, pp. 1--8.

\bibitem{Lent97}
C.~S. Lent and P.~D. Tougaw, ``A device architecture for computing with quantum
  dots,'' vol.~85, no.~4, pp. 541--557, Apr 1997.

\bibitem{Timler02}
J.~Timler and C.~Lent, ``Power gain and dissipation in quantum-dot cellular
  automata,'' vol.~91, no.~2, pp. 823--831, 2002.

\bibitem{Pitters11}
J.~Pitters, L.~Livadaru, M.~Haider, and R.~Wolkow, ``Tunnel coupled dangling
  bond structures on hydrogen terminated silicon surfaces,'' \emph{J. Chemical
  Physics}, vol. 134, no.~6, 2011.

\bibitem{Chaves15}
J.~F. Chaves \emph{et~al.}, ``Towards reversible qca computers: Reversible
  gates and alu,'' Feb 2015, pp. 1--4.

\bibitem{Singh17}
G.~Singh, R.~Sarin, and B.~Raj, ``Design and analysis of area efficient qca
  based reversible logic gates,'' vol.~52, pp. 59 -- 68, 2017.

\bibitem{Mukhopadhyay15}
D.~Mukhopadhyay and P.~Dutta, ``A study on energy optimized 4 dot 2 electron
  two dimensional quantum dot cellular automata logical reversible
  flip-flops,'' vol.~46, no.~6, pp. 519 -- 530, 2015.

\bibitem{Chabi17}
A.~M. Chabi, A.~Roohi, H.~Khademolhosseini, S.~Sheikhfaal, S.~Angizi, K.~Navi,
  and R.~F. DeMara, ``Towards ultra-efficient qca reversible circuits,''
  vol.~49, no. Supplement C, pp. 127 -- 138, 2017.

\bibitem{Sen17}
M.~Goswamia, B.~Sen, R.~Mukherjee, and B.~K. Sikdar, ``Design of testable adder
  in quantumdot cellular automata with fault secure logic,'' vol.~60, no.~C,
  pp. 1--12, Feb. 2017.

\bibitem{Lent06}
C.~S. Lent, M.~Liu, and Y.~Lu, ``Bennett clocking of quantum-dot cellular
  automata and the limits to binary logic scaling,'' \emph{Nanotechnology},
  vol.~17, no.~16, p. 4240, 2006.

\bibitem{Ottavi11}
M.~Ottavi \emph{et~al.}, ``Partially reversible pipelined qca circuits:
  Combining low power with high throughput,'' vol.~10, no.~6, pp. 1383--1393,
  Nov 2011.

\bibitem{Walus06}
K.~Walus and G.~A. Jullien, ``Design tools for an emerging soc technology:
  Quantum-dot cellular automata,'' vol.~94, no.~6, pp. 1225--1244, June 2006.

\bibitem{Bhanja09}
S.~Srivastava \emph{et~al.}, ``Estimation of upper bound of power dissipation
  in qca circuits,'' vol.~8, no.~1, pp. 116--127, Jan 2009.

\bibitem{Sill18}
F.~{Sill Torres}, R.~{Wille}, P.~{Niemann}, and R.~{Drechsler}, ``An
  energy-aware model for the logic synthesis of quantum-dot cellular
  automata,'' vol.~37, no.~12, pp. 3031--3041, Dec 2018.

\bibitem{Huang05}
J.~Huang \emph{et~al.}, ``Tile-based qca design using majority-like logic
  primitives,'' vol.~1, no.~3, pp. 163--185, Oct. 2005.

\bibitem{Walter19a}
M.~Walter, R.~Wille, F.~{Sill Torres}, D.~Gro{\ss}e, and R.~Drechsler,
  ``Scalable design for field-coupled nanocomputing circuits,'' 2019, pp.
  197--202.

\bibitem{Anderson14}
N.~G. Anderson and S.~Bhanja, \emph{Field-coupled Nanocomputing: Paradigms,
  Progress, and Perspectives}, 1st~ed.\hskip 1em plus 0.5em minus 0.4em\relax
  New York: Springer, 2014.

\bibitem{Liu2013design}
W.~Liu, E.~E. Swartzlander~Jr, and M.~O’Neill, \emph{Design of semiconductor
  QCA systems}.\hskip 1em plus 0.5em minus 0.4em\relax Artech House, 2013.

\bibitem{Hennessy01}
K.~Hennessy and C.~Lent, ``Clocking of molecular quantum-dot cellular
  automata,'' vol.~19, no.~5, pp. 1752--1755, 2001.

\bibitem{Rahimi16}
E.~{Rahimi}, ``Energy dissipation of quantum-dot cellular automata logic
  gates,'' \emph{Micro Nano Letters}, vol.~11, no.~7, pp. 369--371, 2016.

\bibitem{Jeanniot16}
N.~Jeanniot, A.~Todri-Sanial, P.~Nouet, G.~Pillonnet, and H.~Fanet,
  ``Investigation of the power-clock network impact on adiabatic logic,'' May
  2016, pp. 1--4.

\bibitem{Bajec12}
I.~L. Bajec and P.~Pečar, ``Two-layer synchronized ternary quantum-dot
  cellular automata wire crossings,'' \emph{Nanoscale Res. Lett.}, vol.~7,
  no.~1, 2012.

\bibitem{Timler03}
J.~Timler and C.~S. Lent, ``Maxwell's demon and quantum-dot cellular
  automata,'' vol.~94, no.~2, pp. 1050--1060, 2003.

\bibitem{zulehner17}
A.~Zulehner and R.~Wille, ``Make it reversible: Efficient embedding of
  non-reversible functions,'' 2017, pp. 458--463.

\bibitem{Huang06}
J.~Huang, X.~Ma, and F.~Lombardi, ``Energy analysis of qca circuits for
  reversible computing,'' in \emph{Nanotechnology}, vol.~1, June 2006, pp.
  39--42.

\bibitem{Tougaw99}
A.~Gin, P.~Tougaw, and S.~Williams, ``An alternative geometry for quantum-dot
  cellular automata,'' \emph{Journal of Applied Physics}, vol.~85, no.~12, pp.
  8281--8286, 1999.

\bibitem{Walter18}
M.~{Walter}, R.~{Wille}, D.~{Große}, F.~S. {Torres}, and R.~{Drechsler}, ``An
  exact method for design exploration of quantum-dot cellular automata,'' March
  2018, pp. 503--508.

\bibitem{Wille19}
R.~{Wille}, M.~{Walter}, F.~{Sill Torres}, D.~{Große}, and R.~{Drechsler},
  ``Ignore clocking constraints: An alternative physical design methodology for
  field-coupled nanotechnologies,'' in \emph{2019 IEEE Computer Society Annual
  Symposium on VLSI (ISVLSI)}, July 2019, pp. 651--656.

\bibitem{Walter19b}
M.~Walter, R.~Wille, D.~Gro{\ss}e, F.~{Sill Torres}, and R.~Drechsler,
  ``Placement \& routing for tile-based field-coupled nanocomputing circuits is
  {NP}-complete,'' vol.~15, no.~29, 2019.

\end{thebibliography}

\end{document}